\documentclass[aps,prb,twocolumn,showpacs]{revtex4}

\begin{document}
\title{Paramagnetic limiting in  ferromagnetic superconductors with triplet pairing}

\author{V. P. Mineev}

\affiliation{Commissariat \`a l'Energie Atomique,
INAC/SPSMS, 38054 Grenoble, France}
\date{\today}

\begin{abstract}
The spin susceptibility in the uranium ferromagnet superconductors is calculated.
 There is shown that the absence of superconductivity paramagnetic limitation for the field directions perpendicular to the direction of the spontaneous magnetization is explained  by the itinerant ferromagnet band splitting rather than by a rotation of magnetization toward the external field direction. The qualitative description of the upper critical field temperature dependence is given.
 
\end{abstract}


\maketitle

\section{Introduction}

It is commonly believed that in the ferromagnet superconducting uranium compounds \cite{Saxena,Aoki01,Huy07} we deal with triplet superconductivity.  In particular, it  is due to the fact that the upper critical field strongly exceeds the  Pauli limit.
However, the paramagnetic limitation of triplet superconductivity is inessential only then  the external field direction is parallel to the spin quantization axis.
On the contrary, it is quite essential for the field orientation perpendicular to the spin quantization direction.\cite{Book,Choi} In actuality the situation is 
the opposite.
In two of uranium superconducting compounds URhGe and UCoGe the upper critical field  in the direction of spontaneous magnetization $\hat c$ is  about  the paramagnetic limiting field in this materials. At the same time 
the upper critical field in the perpendicular to magnetization directions is much higher than the paramagnetic limiting field.\cite{Hardy05,Huy08,Slooten}  Moreover this property persists also in the reentrant superconducting state of the URhGe \cite{Levy05,Levy07}, where the  superconductivity is reappearing under the magnetic field ~12 Tesla  in $b$ crystallography direction causing alignment of magnetization parallel to $b$ axis. The additional field oriented in $a$ crystallography direction does not destroy the superconducting state till to 20 Tesla !  The similar behavior was recently
found in the UCoGe.\cite{Aoki09,deVisser09} 

One can think that magnetization direction always follows
the direction of the external field that prevents the suppression of superconducting state like it is in the superfluid $^3He-A$ and should be in a superconductor with triplet pairing in the absence of spin-orbital coupling fixing the mutual orientation of  spins quantization axis and the crystalline symmetry directions.\cite {Book}  In uranium compounds the
magnetic anisotropy  is quite  strong. \cite{Shick}
As result, in the superconducting URhGe the field oriented parallel to b axis causes
only tiny rotation of the magnetization direction \cite{Levy05}  till to $H_{c2}\approx 1.3$ Tesla more than twicely exceeding the paramagnetic limiting field.\cite{Hardy05} Hence, the rotation of the magnetization cannot be responsible for the absence of the paramagnetic limitations.                                    

Here we investigate theoretically  such a remarkable behaviour of the ferromagnetic superconductors.
There will be given the microscopic derivation of the paramagnetic susceptibility of the ferromagnet superconductors  for the field orientation perpendicular to the direction of the spontaneous magnetization.  The absence of Pauli limitations of superconductivity  is found related with the itinerant ferromagnet band splitting rather than with the magnetization rotation. The latter is also important at higher fields near the metamagnetic or magnetization rotation phase transition. Hence, the critical field in the itinerant ferromagnets can be calculated ignoring the paramagnet limitations. In conclusion we discuss the upper critical field temperature dependence in the uranium compounds in moderate field region.

\section{Spin susceptibility} 

URhGe and UCoGe  are the orthorhombic ferromagnets with spontaneous magnetization oriented along $c$ crystallography axis. At the temperatures below the Curie
temperature and in the absence of magnetic field
the $c$ component of magnetization has a finite value.  The magnetic
field applied along $b$ axis creates the magnetization along its direction but decreases the
magnetization parallel to $c$. Phenomenologically  it is described by means 
the Landau free energy of ferromagnet in
magnetic field\cite{footnote}
\begin{eqnarray}
F=\alpha_{z0}¥(T-T_c)M_{z}¥^{2}¥+ \beta_{z}¥M_{z}¥^{4}¥~~~~~~~~~~~~~~~~~~\nonumber\\
+\alpha_{y}¥ M_{y}¥^{2}¥
+\beta_{yz}¥
M_{z}¥^{2}¥M_{y}¥^{2}¥-M_{y}¥ H.
\label{ea}
\end{eqnarray}
Here the $y,z$ are directions of the spin axes pinned to $(b,c)$
crystallographic directions correspondingly.
The field induced magnetization along b-direction is
\begin{equation}
M_y=\frac{H}{2(\alpha_y+\beta_{xy}M_z^2)}.
\end{equation} 
Substituting this value back in the eqn. (\ref{ea}) we obtain at $\beta_{xy}M_z^2/\alpha_y<1$, that is certainly true  not so far from the Curie temperature,
\begin{equation}
F=\alpha_{0z}¥\left(T-T_c+\frac{\beta_{yz}H^2}{4\alpha_{z0}\alpha_y^2}\right)M_{z}¥^{2}¥+ \beta_{z}.¥M_{z}¥^{4}.
\label{fe}
\end{equation}
Hence, the Curie temperature 
\begin{equation}
T_{Curie}(H)=T_c-\frac{\beta_{yz}H^2}{4\alpha_{z0}\alpha_y^2}
\end{equation} 
is suppressed by the magnetic field oriented along b-axis.
This type of behavior was observed in UCoGe.\cite{Aoki09} 
The magnetization along z-direction is also decreased
\begin{equation}
M_z^2=\frac{\alpha_{z0}(T_c-T)}{2\beta_z}-\frac{\beta_{yz}H^2}{8\alpha_y^2\beta_z}
\end{equation}
 
The field dependence of magnetization components in URHGe has been reported in the paper \cite{Levy05}. 
For superconducting state realizing in the low field region of the phase diagram
the upper critical fiield for the field orientation along b-axis 
does not exceed 1.3 Tesla.\cite{Hardy05}  At this field  the  magnetization in b-direction is at least 10 times smaller than the magnetization along c-direction which  is practically field independent.\cite{Levy05} Hence, the magnetic field
acting on the electron spins in $\hat z$-direction can be taken equal to exchange field
\begin{equation}
{\bf h}=4\pi M_z(H=0)\hat z.
\end{equation}
The field in $\hat y$-direction is
\begin{equation}
{\bf B}=(H+4\pi M_y(H))\hat y.
\end{equation}

In that follows we shall assume that both phenomena ferromagnetism and superconductivity are determined by the 
spin-up and the spin-down electrons filling two separate bands split by the exchange field $h\sim T_c/\mu_B$.  
Then the magnetic moment of the itinerant electron subsystem is given by
\begin{equation}
{\bf M}=\mu_B~T\sum_n\int\frac{d^3k}{(2\pi\hbar)^3}Tr\mbox{\boldmath $\sigma$}\hat G.
\end{equation}
Here $\mbox{\boldmath $\sigma$}=(\sigma_x,\sigma_y,\sigma_z)$ are Pauli matices.

In the normal state  the Green function in linear  in respect to $B$ approximation is
\begin{equation}
\hat G=\hat G_n-\mu_BB\hat G_n\sigma_y\hat G_n,
\end{equation}
where
\begin{equation}
\hat G_n=\left( \begin{array}{cc}G_{n+}& 0\\ 
0& G_{n-}
\end{array}\right ),~~~~~
G_{n\pm}=\frac{1}{i\omega_n-\xi_{{\bf k}}\pm\mu_Bh}.
\end{equation}
We obtain
\begin{eqnarray}
{\bf M}=\mu_B~T\sum_n\int\frac{d^3k}{(2\pi\hbar)^3}
[\hat z(G_{n+}-G_{n-})~~~~~~~~~\nonumber\\-2\mu_BB\hat yG_{n+}G_{n-}].
\label{Mn}
\end{eqnarray}
For a finite value of the exchange field this is equal to 
\begin{equation}
{\bf M}=\mu_B(N_\uparrow-N_\downarrow)\frac{{\bf h}+{\bf B}}{h}.
\label{Mnn}
\end{equation}
Here $N_{\uparrow,\downarrow}$ are the numbers of electrons in the spin-up and spin-down band.
The corresponding susceptibility is
\begin{equation}
\chi_{yy}=\mu_B(N_\uparrow-N_\downarrow)/h
\end{equation}
On the other hand in absence of the band splitting  that is at $h=0$ the magnetic moment is
\begin{equation}
{\bf M}=2\mu_B^2N_0{\bf B},
\end{equation}
where $N_0$ is the density of states per one electron spin projection.  The susceptibility is given by  the Pauli formula
\begin{equation}
\chi_{yy}(h=0)=2\mu_B^2N_0.
\end{equation}

The superconducting state in  two band itinerant ferromagnet is built
 of pairing states formed either  by spin-up electrons from one band or by spin-down electrons from another band.\cite{Cham,Min04,Min09,Min10} This state is characterized by  two component order parameter
\begin{equation}
\hat \Delta=\left( \begin{array}{cc}\Delta_{{\bf k}\uparrow}& 0\\ 
0& \Delta_{{\bf k}\downarrow}
\end{array}\right ).
\end{equation}
Then instead eqn. (\ref{Mn}) we obtain
\begin{eqnarray}
{\bf M}=\mu_B~T\sum_n\int\frac{d^3k}{(2\pi\hbar)^3}[\hat z(G_{s+}-G_{s-})~~~~~~~~\nonumber\\
-\mu_BB\hat y(G_{s+}G_{s-}+G_{s-}G_{s+}+F_+F_-^\dagger+F_-F_+^\dagger)],
\label{Ms}
\end{eqnarray}
where
\begin{equation}
G_{s\pm}=\frac{-i\omega_n-\xi_{{\bf k}\pm}}{\omega_n^2+\xi_{{\bf k}\pm}^2+|\Delta_{\uparrow,\downarrow}|^2},~~~~F_{\pm}=\frac{\Delta_{\uparrow,\downarrow}}{\omega_n^2+\xi_{{\bf k}\pm}^2+|\Delta_{\uparrow,\downarrow}|^2}
\end{equation}
are the superconducting state Green functions and\\ $\xi_{{\bf k}\pm}=\xi_{{\bf k}}\mp\mu_Bh$.

The straightforward calculation shows, that at the band splitting exceeding the superconducting gaps
$\mu_Bh\gg|\Delta_\pm|$, even at $T=0$,
\begin{eqnarray}
\frac{T\sum_n\int d\xi~[2G_{n+}G_{n-}-2
G_{s+}G_{s-}-F_+F_-^\dagger-F_-F_+^\dagger]}
{T\sum_n\int d\xi~G_{n+}G_{n-}}\nonumber\\
\sim\sum_{\alpha\beta=\uparrow,\downarrow}\frac{\Delta_{{\bf k}\alpha}\Delta_{{\bf k}\beta}^*}{(\mu_Bh)^2}\ln\frac{(\mu_Bh)^2}{\Delta_{{\bf k}\alpha}\Delta_{{\bf k}\beta}
}\ll 1~~~~~~~~~~~~~~~~~
\end{eqnarray}
It implies that the susceptibility in the superconducting state practically keeps its normal state value.\cite{Sam} The paramagnetic limiting field formally proves to be of the order of the exchange field
\begin{equation}
H_p\approx \frac{h}{\ln(\mu_Bh/|\Delta|)}.
\end{equation}
Hence, so long the band splitting is larger than the gap, the paramagnetic suppression of the superconducting state by the field perpendicular to the spontaneous magnetization is absent.

On the contrary at $h=0$ the formal calculation from the equation(\ref{Ms}) yields the susceptibility
\begin{equation}
\chi_{yy}(h=0, T)=2\mu_B^2N_0\int\frac{d\Omega}{4\pi}Y(\hat{\bf k},T),
\end{equation}
where 
$$
Y(\hat{\bf k},T)=\frac{1}{4T}\int_{-\infty}^{+\infty}\frac{d\xi}{\cosh^2(\sqrt{\xi^2+\Delta_{\bf k}^2}/2T)}
$$
is generalized Yosida function. The susceptibility  $\chi_{yy}(h=0, T)$  tends to zero at $T\to 0$.\cite{Book}
Thus, the magnetic field directed perpendicular to the Cooper pairs spins in a nonferromagnet  superconductor with triplet pairing suppress superconductivity like it does in the usual superconductors with singlet pairing.

All the formulated conclusions are valid at moderate magnetic fields when $M_y(H)\ll M_z(H)$. At external fields 
 of the order of exchange field $H\sim h$, the equilibrium magnetzation align itself parallel to the external field. In this conditions the paramagnetic limitation of superconductivity is absent as well.
 
\section{Concluding remarks}

In general there are three mechanisms of the magnetic field influence on the superconducting state in the superconductors with triplet pairing\cite{Luk}: (i) the orbital depairing, (ii) paramagnetic limiting, and (iii) stimulation or suppression of nonunitary superconductivity due to magnetic field dependence 
of density of states \cite{Amb}.

We have demonstrated here that the superconducting state in the itinerant superconductors with triplet pairing is not a subject of the paramagnetic limiting. For the completeness let us briefly look at the two other source of the field influence.

Making use eqn. (\ref{Ms}) one can show that the superconducting spontaneous magnetization
\begin{equation}
{\bf M}_s=\hat z \mu_B\left [N_\uparrow-N_\downarrow+(N_{0\uparrow}^\prime |\Delta_\uparrow|^2-
N_{0\downarrow}^\prime |\Delta_\downarrow|^2)\ln\frac{\varepsilon_F}{T_s}\right ].
\label{Mns}
\end{equation}
is slightly modified in comparison with its normal state value 
\begin{equation}
{\bf M}_n=\hat z \mu_B\left [N_\uparrow-N_\downarrow
\right ].
\label{Mnnn}
\end{equation}
Here, $N_{0\uparrow}^\prime$ and $N_{0\downarrow}^\prime$ are the energy derivatives of the density of states at the Fermi level for the spin up and spin down band correspondingly.  The spontaneous magnetization change causes the corresponding energy shift under magnetic field $H_z$ parallel to $\hat z$ direction that  leads in its turn to the  criticall temperature shift. To avoid the cumbersome formula we write it for the case of presence only the spin-up pairing 
\begin{equation}
\frac{\delta T_s}{T_s}=\mu_BH_z\frac{N_{0\uparrow}^\prime}{N_{0\uparrow}}\ln\frac{\varepsilon_F}{T_s} 
\label{deltaT}
\end{equation}

On the other hand the magnetic field $\bf H$ directed perpendicular to the direction of spontaneous magnetization does not cause a linear in $H$ shift in the free energy of superconducting state. Hence, for this field direction the third mechanism  of magnetic field influence on the superconducting state is also ineffective. 
The statement is valid in the moderate  fields when the inequality $M_y(H)\ll M_z(H)$ takes place.

Thus the only orbital mechanism suppression of  superconductivity is essential. 
Experimentally , in UCoGe for the field directed perpendicular to the spontaneous magnetization
there was observed the pronounced upper critical field upward curvature\cite{Huy08,Slooten,Aoki09}
apparently  related with the magnetic field dependence of the effective mass in this material 
\cite{Aoki09}. Indeed,  for an
  orthorhombic superconductor under magnetic field directed along $b$ direction \cite{footnote2}  
the  Ginzburg-Landau formula for the critical temperature ( for simplicity we limit ourself by the one band case) is
\begin{equation}
T_s(H)=T_{s0}\left (1-C\frac{H}{m_a^*(H)m_c^*(B)}\right ),
\label{Tc}
\end{equation} 
where $C$ is a  constant  with dimensionality $m^2/H$. We see, that the decreasing of the effective mass with increasing of magnetic field followed by saturation of its field dependence found at moderate fields in the paper
 \cite{Aoki09} inevitably causes the appearance of the upward curvature in 
magnetic field dependence of the critical temperature as well of the upper critical field.

\acknowledgments

This work was partly supported by the grant SINUS of  Agence Nationale de la Recherche.

The author is indebted to J.-P. Brison for the enlightening discussion and the interest to the paper.


\begin{thebibliography}{99}
\bibitem{footnote2} Looking for the orbital suppression of superconductivity by the magnetic field directed along $b$ axis one can neglect by the action of spontaneous magnetization directed along $c$ axis on the electron charges. The reason for this is that the latter magnetic field can be estimated as 
$\mu_U/V_{ec}$ where $\mu_U$ is magnetic moment per Uranium atom and $V_{ec}$ is the elementary 
cell volume. For the URhGe this is of the order 100 Gauss  and it is even less in the case of UCoGe.   The scale of the measured upper critical fields  noticeably exceeds such a small value.
\end{thebibliography}
\end{document}